# Generative-AI and the transformation of workforce. A job postings-driven analysis

Diana Maria POPA, Simona-Vasilica OPREA*, Adela BÂRA
*Department of Economic Informatics and Cybernetics, Bucharest University of Economic Studies, Bucharest, Romania*
*Corresponding author. E-mail: *simona.oprea@csie.ase.ro*

**Abstract** This paper investigates how generative-artificial intelligence (AI) is reshaping job requirements, skill compositions and sectoral dynamics across global labor markets. It examines the evolving frequency and framing of AI-related competencies in job postings, exploring whether generative-AI functions primarily as an augmentative or substitutive force in the workplace. A large-scale, multi-source corpus of over 150,000 English-language job postings (2018-2025) is compiled from twelve open-access datasets and one public API. The analytical framework integrates lexical skill extraction, semantic framing, topic modeling (BERTopic, LDA, KMeans) and time-series forecasting (ARIMA). Skill mentions are categorized into five dimensions: AI_Data, Routine, Soft_Meta, Domain_Specific and Leadership, while cross-sectoral analyses and correlation matrices quantify interdependencies between competencies. Sentence-transformer embeddings and cosine similarity are used to compute a Framing Index, distinguishing augmentation- versus automation-oriented discourse. Investigating job postings, our research contributes a replicable, data-driven methodology for mapping the diffusion of AI-related skills across industries and time. Results reveal a sharp post-2021 increase in AI-related skill mentions: "prompt engineering", "fine-tuning" and "model validation", accompanied by a decline in routine tasks: "data entry" and "manual coding". Forecasts suggest sustained growth in AI_Data and Soft_Meta skills through 2025, signaling a structural convergence toward hybrid human-AI expertise as a new foundation of employability.

**Keywords:** generative-AI; labor market transformation; skill evolution; semantic framing; topic modeling; workforce adaptation

## 1. Introduction

The rise of generative artificial intelligence (AI) has triggered deep and multifaceted transformations in labor market structures and the skillsets required across occupations. Unlike traditional automation, which primarily targeted repetitive physical or computational tasks, generative-AI systems such as ChatGPT, Copilot and Claude can produce text, images, code, abstraction and analytical insights with human-like fluency and machine-like rigor [1], [2]. This capability fundamentally redefines the nature of work, creativity and professional identity across some industries. Consequently, job requirements are undergoing rapid transformation, demanding new technical and cognitive competencies. Traditional job structures, once centered around the execution of defined tasks, are evolving into collaborative ecosystems where humans and AI systems co-create value. This transformation affects the skills required for specific professions and also the broader dynamics of organizational culture, education and regulations [3]. Thus, understanding how generative-AI influences job requirements and what forms of workforce adaptation are necessary has become an important question for researchers, policymakers and business leaders alike [4].

The changing nature of job requirements can be observed in several interrelated dimensions [5], [6]. First, *AI literacy and human-AI collaboration* are emerging as core competencies across all sectors. Workers are expected to know how to collaborate with generative models rather than merely operate them. Skills such as prompt engineering, model interpretation, fine-tuning and evaluation of AI outputs are increasingly valued. Second, *creative and analytical roles* are being redefined; professionals in fields like journalism, marketing and software development are shifting from producing content to curating, supervising and strategically integrating AI-generated content [7]. Finally, generative-AI has led to the emergence of entirely *new job roles*, from prompt engineers to AI ethicists, reshaping the labor market around hybrid human-machine expertise [8].

One of the most pressing challenges accompanying the rapid expansion of generative-AI is the widening gap between existing educational frameworks and the evolving demands of the labor market [9]. As technological innovation accelerates, academic curricula, vocational training and certification standards struggle to keep pace, resulting in a growing skills adaptation deficit. In many cases, educational systems continue to prepare graduates for roles that no longer reflect the realities of an AI-driven economy [10]. From a technical and institutional perspective, this misalignment manifests in two main ways. First, conventional training pathways remain anchored in pre-AI paradigms that emphasize procedural knowledge and repetitive execution. Such an approach fails to equip learners with essential skills. This imbalance produces an oversupply of workers with outdated technical competencies and an undersupply of individuals capable of prompt design, multimodal analysis and uncertainty management. Second, the lack of unified frameworks for AI literacy and competency assessment slows the diffusion of interdisciplinary expertise across sectors such as healthcare [11], law and engineering.

Given these transformations, *workforce adaptation* has become an economic and social necessity. The rapid pace of AI innovation means that skills have a shorter lifespan, making continuous reskilling and lifelong learning essential for employability. Educational institutions and organizations have to redesign training programs to incorporate AI-related content, while individuals cultivate adaptability and self-directed learning habits. Simultaneously, human-centric competencies such as reasoning, creativity and interdisciplinary collaboration are becoming more valuable, as they represent areas where human intelligence complements AI capabilities rather than competes with them [12]. At the organizational level, adaptation involves restructuring workflows, redefining job roles and fostering an innovation-oriented culture that embraces AI as a collaborative partner. By automating routine processes and augmenting creative and analytical functions, it enables professionals to focus on higher-value, strategic and innovative tasks. The future of work will therefore be characterized by symbiosis, where humans and AI systems learn from one another, co-evolve and jointly drive progress [13], [14].

Our research explores this transformation, analyzing how generative-AI is reshaping job requirements, identifying emerging skills and professions, and outlining the pathways for effective workforce adaptation in the era of intelligent automation. We investigate how the emergence and diffusion of generative-AI technologies reshape the composition of required skills in job postings. Job advertisements provide a high-resolution and near-real time representation of employers' expectations, revealing both technical and transversal competencies. By systematically processing these texts, we detect sectoral and temporal transformations in labor demand. Finally, we aim to answer the following research questions (RQ):

*RQ1:* How has mention-frequency of AI-related skill tokens (like "prompt engineering", "model validation") changed across occupations and sectors in time?

*RQ2:* Are generative-AI skill mentions correlated with reductions in mentions of routine tasks (like "data entry", "rote coding") within the same job titles?

*RQ3:* How do the semantic frames surrounding AI (augmentation vs. automation) evolve across sectors, and what do these shifts indicate about the perceived role of AI in the workplace?

## 2. Literature review

The diffusion of generative-AI and digital technologies has intensified transformations in global labor markets, reshaping both the structure of work and the demand for skills. A growing body of empirical research highlights how globalization, digitization and automation jointly drive shifts in occupational profiles, productivity and workforce adaptability. Recent studies employ diverse methodologies, from large-scale analyses of job postings to firm-level field experiments, to capture the evolving interplay between human capabilities and AI systems. This emerging literature provides a multidimensional understanding of how AI adoption influences skill demand, job design and productivity across sectors, forming the foundation upon which the current research builds.

For instance, [15] examines how globalization, digitization and rapid technological change are reshaping labor market skill requirements, emphasizing the need for continuous workforce adaptation. Using over 500,000 Lithuanian job postings, the authors applied natural language processing (NLP) and clustering techniques to automatically identify emerging skill profiles and job categories. Job requirements were extracted using regex-based parsing, while BERT sentence transformers were employed for feature vectorization. The resulting clusters were used to generate job descriptions with generative-AI, which experts confirmed as accurate and practical. Overall, the findings demonstrate that NLP and clustering can enable real-time monitoring of labor market skill demand, supporting data-driven workforce and education policy planning. Also, [16] show in a large firm-scale deployment that a generative-AI assistant increased customer-support productivity by ~15%, with the biggest gains for lower-tenure agents, evidence for complementarity and skill-gap compression. [17] likewise find faster task completion and higher writing quality with ChatGPT in randomized experiments. Further, [18] meta-analyze >100 studies and clarify when human-AI teams outperform: content-creation benefits are stronger than for decision tasks. Together, these studies support the "augmentation" side as other researchers [19]. Using administrative claims and occupation exposure metrics, [20] show AI exposure predicts unemployment risk for specific occupation-region cells, highlighting uneven impacts. At the vacancy level, [21] document rapid growth in AI-related postings and reconfiguration of skills within adopting establishments, consistent with our postings-based approach.

Analyzing millions of postings, [22] finds a marked rise in AI skill terms and a drift toward skill-based hiring (with reduced degree requirements), empirical backing for our RQ1 on AI-token frequencies. Thus, our lexicon/taxonomy pipeline aligns with this stream by operationalizing category-level counts over time. A management-practice lens shows nuance: [23] synthesize six paradoxes of generative-AI customer service (e.g., efficiency vs. empathy), reinforcing why augmentation narratives dominate in service contexts even as automation pressures rise. In research workflows, [24] document algorithmic management in science, again indicating hybrid human-AI orchestration rather than wholesale substitution. In marketing, [25] outline how generative-AI restructures content creation, product development and customer interaction, consistent with our sectoral spikes post-2021 and other researchers in different sectoral anchor [26]. For creative production, [27] show text-to-image generative-AI boosts creative productivity and value signals at scale. In healthcare, [28] (systematic review) and [29] (meta-analysis) temper enthusiasm: governance, bias and mixed diagnostic accuracy argue for cautious augmentation, echoing our legal/health findings.

Additionally, [30] show "double-edged" generative-AI roles in creative digital work, gains with new coordination costs, while [31] theorize generative-AI's transformative mechanisms on digital platforms (e.g., recombination, autonomy), offering constructs one can reuse for "augmentation vs. automation" framing at the workflow level.

Across industry contexts, [32] review Large Language Models (LLMs) industrial applications; [33] link AI adoption to worker well-being dynamics, reminding that how AI is integrated matters for job quality; [34] synthesize organizational productivity effects and sustainability implications; [35] quantify skill-bias patterns in labor demand under AI. These aspects further complement our cross-sector trend analysis.

While the aforementioned research delivers important insights, three gaps remain: (1) existing work seldom uses job-posting texts at large scale to track skill-mention frequencies across time and sectors; (2) few studies operationalize both skill-demand evolution and semantic-framing shifts (augment vs. automate) in the same dataset; (3) there is limited longitudinal, multi-sector evidence covering the diffusion of generative (rather than generic) AI skills from ~2018 onward. Our research addresses these gaps by combining a multi-source job-posting corpus over 2018-2025, lexicon-based skill taxonomy, semantic-framing analysis and forecasting of skill trajectories across sectors.

## 3. Methodology

### 3.1 Input data

The empirical foundation of this research consists of a large-scale, multi-source corpus of job postings, aggregated from 12 publicly available open-access datasets and one public API (presented in Table 1). The datasets are selected for their transparent licensing, reproducibility and global representativeness across industries and regions. All data sources are open, either hosted on Kaggle or accessible through the ReliefWeb public API, making the entire corpus suitable for academic and non-commercial use. The data collection strategy is designed to achieve both longitudinal and cross-sectoral coverage. To this end, job postings are gathered from major employment platforms such as Indeed, CareerBuilder, SimplyHired, Reed UK and JobStreet, complemented by LinkedIn job datasets focusing on AI, data science and business analytics roles. In addition, domain-specific sources, such as *eMedCareers* (healthcare) and *ReliefWeb* (digital development and humanitarian work) are integrated to ensure balanced representation of both traditional and emerging fields.

All datasets include structured metadata (like job title, description, publication date, location), but only two fields are retained for the semantic and temporal analyses: the job description text and the publication date. This ensures analytical consistency across diverse file formats (CSV, XML, JSON, LDJSON). The preprocessing pipeline harmonizes the input from multiple sources into a unified schema, removing duplicates, standardizing encodings and cleaning boilerplate or repetitive content. For the Kaggle datasets, each source was downloaded manually and stored locally in standardized formats (CSV, XML, LDJSON). For the ReliefWeb API, a Python script was developed to collect international job postings between 2018 and 2025 directly from the official REST endpoint.

Table 1. Input data sources

| No. | Dataset Name | Description | Records (approx.) | Format | Access Link | Platform | Open Access |
|---|---|---|---|---|---|---|---|
| 1 | ReliefWeb API (2018–2025) | International postings in digital development, public administration, and humanitarian work | 50,000+ | JSON | ReliefWeb API | ReliefWeb | Yes |
| 2 | eMedCareers EU Job Board | Healthcare postings from EU markets | 40,000 | XML | Healthcare Jobs EU | Kaggle | Yes |
| 3 | Indeed Job Postings Dataset | General global job listings | 100,000+ | CSV | Indeed Dataset | Kaggle | Yes |
| 4 | CareerBuilder Job Listing (USA) | U.S.-based job postings | 20,000 | LDJSON | CareerBuilder Dataset | Kaggle | Yes |
| 5 | SimplyHired Job Listing (2019) | General U.S. job market sample | 10,000 | XML | SimplyHired Jobs | Kaggle | Yes |
| 6 | Reed UK Job Board | UK employment data | 50,000 | CSV | Reed UK Dataset | Kaggle | Yes |
| 7 | LinkedIn Data Analyst Jobs (USA, Canada, Africa) | Regional data analytics and AI-related roles | 30,000 | CSV | LinkedIn Analyst Jobs | Kaggle | Yes |
| 8 | JobStreet All Job Dataset | Southeast Asian job postings | 60,000 | CSV | JobStreet Dataset | Kaggle | Yes |
| 9 | AI Job Dataset (Global) | AI/ML-related job listings with required skills | 25,000 | CSV | Global AI Job Dataset | Kaggle | Yes |
| 10 | AI & ML Job Listings (LinkedIn) | U.S. LinkedIn postings related to AI and ML | 15,000 | CSV | AI & ML Listings | Kaggle | Yes |
| 11 | Business Analyst Job Listings (LinkedIn) | Business and analytics roles, 2024 | 10,000 | CSV | Business Analyst Jobs | Kaggle | Yes |
| 12 | AI Job Dataset | Alternative version including additional skill annotations | 15,000 | CSV | AI Job Dataset v2 | Kaggle | Yes |

The obtained dataset contained job postings, each structured into two primary variables: Date-representing the posting's publication date, enabling longitudinal trend analysis from 2018 to 2025; and Description-the complete text of each job advertisement, used as input for NLP and semantic modeling tasks.

### 3.2 Framework

The proposed analytical pipeline developed for this research operationalizes the conceptual framework outlined through a sequence of Python-based modules (as presented in Figure 1), designed to capture the evolving dynamics of skill requirements across multiple industries. Each stage corresponds to a distinct methodological layer, from data acquisition and preprocessing to semantic modeling and forecasting, ensuring both reproducibility and interpretability.

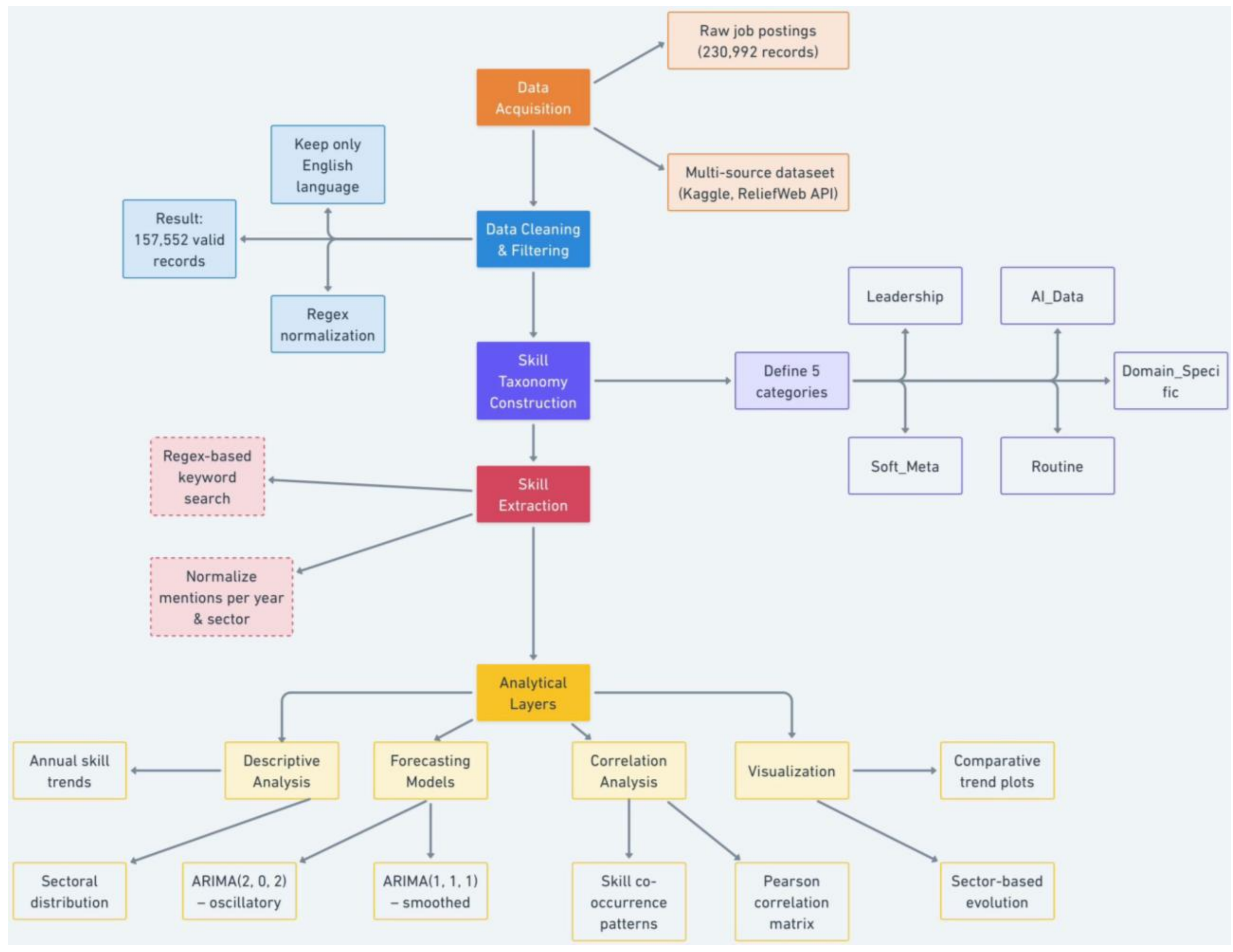

Figure 1. Full analytical pipeline

**3.2.1. Text preprocessing and structuring**

The analytical framework is implemented as a modular pipeline designed to transform the heterogeneous corpus of job descriptions into a structured, analyzable format suitable for semantic modeling, skill extraction and temporal analysis. The records in the final dataset contain two essential variables: a publication date, ranging from 2018 to 2025, and a textual description representing the body of the job posting.

**3.2.1.1. Data cleaning and standardization**

The initial phase of the framework consists of cleaning and standardizing the raw corpus of job descriptions. This pipeline combines language detection, text normalization and heuristic filtering to retain only meaningful English-language content.

Textual cleaning relies entirely on rule-based regular expressions to remove control characters, redundant whitespace and recurring non-informative patterns typically found in recruitment templates. Standardized expressions such as “Equal Opportunity Employer”, “About the Company” and “Benefits include” are eliminated to preserve only the substantive portions of the text related to job content and required competencies.

A language detection module is integrated to automatically exclude non-English records, thereby maintaining linguistic consistency across the dataset. The publication year is derived from the date field to enable longitudinal comparisons, while extremely short or incomplete entries are discarded.

Following this multistage cleaning and filtering process, the dataset is reduced to 157,552 (out of an initial 230,922) valid English-language postings, each containing relevant descriptions suitable for semantic and temporal analysis. This procedure ensures that the subsequent modeling stages are grounded in a reliable and methodologically transparent textual foundation.

#### 3.2.1.2. Semantic anchors and skill taxonomy construction

A central methodological element of the analytical framework is the creation of two complementary lexical resources: a) a set of semantic anchors for framing analysis, and b) a hierarchical skill taxonomy for classification. Both are designed to capture the linguistic and conceptual evolution of generative-AI discourse across professional sectors.

The semantic anchors are structured into three main vectors: *AI*, *Augment* and *Automate*, representing distinct orientations in the human-machine narrative. The *AI* anchors include tokens such as generative-AI, LLMs, prompt engineering and model monitoring, reflecting the core technical lexicon of the field. The *Augment* group comprises expressions like assist, co-create, hybrid intelligence and human-in-the-loop, emphasizing human-AI collaboration. In contrast, the *Automate* anchors capture substitutional language, including automated, replace and robotic process automation. This structuring is performed manually, following a qualitative analysis of the job descriptions and their linguistic patterns. Each anchor is selected based on its frequency, contextual meaning and relevance to one of the three conceptual categories (AI, Augment, Automate). To ensure contextual precision, domain-specific subsets are also defined for Legal, Healthcare, Software Engineering and Design, enabling sector-based filtering of job postings via lexical triggers such as contract review, diagnosis or creative direction. These subsets are also manually curated after analyzing the typical terminology and task descriptions specific to each field, so that only the most representative and context-appropriate expressions are retained for sector-level filtering.

Complementing this, the skill taxonomy groups extracted terms into *five standardized macro-categories*: AI_Data, Routine, Soft_Meta, Domain_Specific and Leadership (as in Table 2). Each category is constructed based on conceptual reasoning and iterative inspection of representative job descriptions. For example, AI_Data aggregates explicit AI and machine learning skills (GPT, fine-tuning, MLOps), Routine referres to procedural tasks (data entry, manual coding), while Soft_Meta and Leadership capture transversal and managerial competences (strategic planning, communication).

Table 2. Skills classified into 5 dimensions

| Category | Description |
| --- | --- |
| **AI_Data** | AI, ML and data-centric technical skills |
| **Routine** | repetitive, procedural or clerical tasks |
| **Soft_Meta** | transversal analytical, ethical and communication skills |
| **Domain_Specific** | sector-specific knowledge and applied expertise |
| **Leadership** | strategic, managerial and decision-making abilities |

### 3.2.2. Skill extraction and categorization

Skill detection is implemented through a lexicon-based approach operationalized in the next Python module. This procedure relies on a curated hierarchical dictionary stored in JSON format, integrating canonical skill taxonomies with additional generative-AI terms manually compiled (for instance, prompt engineering, fine-tuning, model monitoring). The detection process used regular expressions to identify skill mentions and applied pattern normalization to capture linguistic variations such as “Python programming” or “proficiency in Python”. Each skill category, AI_Data, Routine, Soft_Meta, Domain_Specific and Leadership, is defined by a set of regular expression patterns automatically compiled and applied to all cleaned job descriptions.

For every posting, Boolean indicators are generated to mark the presence or absence of terms belonging to each skill category. Annual aggregates are then computed, yielding normalized frequencies expressed as mentions per 1,000 postings. These results are exported in a file and visualized as a graphic, representing temporal dynamics of core skill categories across 2018-2025. This lexicon-driven framework offers high interpretability and consistency across heterogeneous datasets, while remaining adaptable to emerging terminology. The proposed method captures the growing diffusion of AI and data-related competencies in multiple sectors, including e-commerce, where AI-driven personalization, marketing automation and data analytics have become dominant skill clusters.

### 3.2.3. Comparative semantic and framing analysis

#### 3.2.3.1. Semantic-shift and framing analysis

To examine how AI is linguistically and conceptually represented within job postings, a semantic framing analysis is conducted. The objective is to determine whether AI is portrayed primarily as a tool for augmentation (supporting and enhancing human work) or automation (replacing human labor). Three semantic anchor groups are defined (as in Table 3) and stored in a JSON configuration file to guide the analysis.

Table 3. Semantic anchor groups

| Semantic anchor | Description |
|---|---|
| **AI anchors** | capturing technological entities such as GPT, deep learning and transformer model |
| **Augment anchors** | emphasizing collaboration and assistance, including assist, decision support and human-in-the-loop |
| **Automate anchors** | associated with full replacement of human input, such as automation, autonomous and replace |

Each job description is converted into dense numerical embeddings using the SentenceTransformer model *all-MiniLM-L6-v2*. The script then computes cosine similarities between each job vector and the three anchor vectors, obtained by averaging embeddings of the respective term lists. The Framing Index (FI) is defined as:

$$FI = similarity(Augment) - similarity(Automate) \quad (1)$$

This continuous score quantifies whether the discourse surrounding AI framed it as a collaborative enabler (positive FI) or as a replacement technology (negative FI). The process is fully implemented in Python using the *sentence-transformers* and *pandas* libraries, with outputs including yearly averages and graphical visualizations of similarity trends.

#### 3.2.3.2. Cross-sectoral framing trends

Job descriptions are further categorized into nine macro-sectors (IT, Healthcare, Legal, Education, Finance, Design, Logistics, Management and Sales) using keyword-based classification rules embedded in the analysis scripts. Aggregating results by year and sector yields a longitudinal dataset covering the 2018-2025 interval. The classification rules are defined manually through an iterative process of lexical inspection, in which sector-specific terminology is identified and validated based on its frequency and contextual relevance within job descriptions. Each category includes a controlled vocabulary of professional and technical terms ("developer", "nurse", "lawyer", "designer", "manager") that serve as regular expression triggers for sector detection. These rules are implemented programmatically in Python, using precompiled regex patterns to ensure consistent and reproducible tagging of over 150,000 job postings.

The resulting classification enables descriptive statistics and also temporal and cross-sectoral analyses of skill dynamics, providing the empirical foundation for the longitudinal trends discussed in the results section.

### 3.2.4. Semantic topic modeling with BERTopic, LDA and KMeans

Three topic modeling techniques are applied in order to extract relevant topics from job postings. A short description of these techniques is provided in Table 4.

Table 4. Brief comparison of the three topic modeling techniques

| Aspect | BERTopic | LDA | KMeans |
|---|---|---|---|
| Core method | Transformer-based contextual embeddings + UMAP + density clustering | Probabilistic topic modeling based on word co-occurrence | Vector-based clustering using semantic embeddings |
| Data input | Sentence embeddings (SentenceTransformer all-MiniLM-L6-v2) | CountVectorizer (lexical frequencies) | Same BERT embeddings as BERTopic |
| No. of topics / clusters | 6 major topics (with temporal tracking) | 6 latent topics | 6 semantic clusters |
| Interpretability | High-semantic, contextual and temporal | Moderate – lexical but clear word patterns | High – semantically coherent, less lexical |
| Representative themes | AI Governance, Healthcare, E-commerce Automation, Creative Design | Customer Service, Healthcare/Education, Data Infrastructure, Marketing/Sales | E-commerce Marketing, Data Analytics, Leadership/Innovation |
| E-commerce focus | Strongly present; AI-driven personalization and automation | Present; customer/service/marketing/sales topic | Prominent – campaign, performance, personalization |

**3.2.4.1. BERTopic analysis**

The first stage employs BERTopic, a transformer-based topic modeling technique designed to capture contextual relationships between words beyond surface lexical patterns. The implementation combines UMAP dimensionality reduction and density-based clustering, allowing the discovery of semantically coherent topic groups directly from the embeddings. Parameters are optimized for interpretability, setting a minimum topic size of 200 and cosine similarity as the distance metric.

Formally, each document $d_i$ is encoded into a contextual embedding vector $e_i$ using a pre-trained transformer model:

$$e_i = f_\theta(d_i) \in \mathbb{R}^n \quad (2)$$

Pairwise similarity between two documents is computed as the cosine of their embedding vectors:

$$\text{sim}(d_i, d_j) = \frac{e_i \cdot e_j}{\|e_i\| \, \|e_j\|} \quad (3)$$

These embeddings are projected into a lower-dimensional manifold using Uniform Manifold Approximation and Projection (UMAP), which preserves both global and local semantic structures:

$$\mathrm{z_i} = \text{UMAP}(e_i) \in \mathbb{R}^k, k < n \quad (4)$$

Finally, the reduced embeddings are clustered using the HDBSCAN algorithm, which groups documents according to their local density in semantic space:

$$C = HDBSCAN(\{z_i\}, min_cluster_size = 200) \quad (5)$$

This method operates by grouping documents according to the proximity of their contextual embeddings rather than by word frequency. As a result, BERTopic captures subtle conceptual distinctions, such as differences between “AI-assisted design” and “AI automation”, that traditional lexical models overlook. The output included document-level topic assignments, topic summaries and temporal dynamics. The longitudinal data produced by BERTopic are exported, and interactive visualizations are generated, allowing the observation of how specific topics (healthcare, finance, e-commerce) evolved in prominence over time. The resulting temporal matrices express the relative topic weight in year topic, normalized by the total number of documents in that interval:

$$T_{y,t} = \frac{N_{y,t}}{\sum_t N_{y,t}} \quad (6)$$

**3.2.4.2. Latent Dirichlet Allocation (LDA) analysis**

To complement the transformer-based modeling, an LDA analysis is conducted to extract probabilistic topics based on lexical co-occurrence patterns. Using a CountVectorizer, the text corpus is converted into a document-term matrix, with English stopwords removed and thresholds applied to retain terms with appropriate frequency. The model is configured to extract six optimal latent topics, aligning with the number of clusters used in the other models.

Formally, LDA assumes that each document is generated as a mixture of topics, and each topic is a distribution over words. The generative process can be summarized as follows:

For each topic $k \in \{1, \dots, K\}$, draw a word distribution:

$$\phi_k \sim \text{Dirichlet}(\beta) \tag{7}$$

Topic-word distribution defines the probability of each word belonging to topic *k*.

For each document $d_i$, draw a topic distribution:

$$\theta_i \sim \text{Dirichlet}(\alpha) \tag{8}$$

Document-topic distribution represents how much each topic contributes to document *i*.

For each word in document:

- Draw a topic assignment:

$$z_{i,j} \sim Multinomial(\theta_i) \tag{9}$$

- Draw a word from the corresponding topic:

$$w_{i,j} \sim Multinomial(\phi_k z_{i,j}) \tag{10}$$

The joint probability of the corpus is then given by:

$$p(W, Z, \Theta, \Phi \mid \alpha, \beta) = \prod_{k=1}^{K} p(\phi_k \mid \beta) \prod_{i=1}^{D} p(\theta_i \mid \alpha) \prod_{j=1}^{N_i} p(z_{i,j} \mid \theta_i)\, p(w_{i,j} \mid \phi_k z_{i,j}) \tag{11}$$

Joint generative probability represents how all documents, topics and words jointly contribute to the model's probabilistic structure.

LDA conceptualizes each job posting as a mixture of underlying topics, each represented by a probability distribution over words. This method's advantage lies in its ability to reveal dominant lexical structures and shared terminology across documents. Although LDA lacks contextual sensitivity compared to embedding-based models, it offers high interpretability through direct inspection of the top-weighted terms per topic. These terms can be visualized as word clouds for qualitative validation of topic coherence. The outputs, including topic-term distributions and document-topic probabilities, are stored for comparative analysis against BERTopic and KMeans. Methodologically, the LDA model served as a lexical benchmark, revealing how frequently co-occurring terms organize thematic domains in traditional probabilistic space.

**3.2.4.3. KMeans clustering analysis**

The third analytical approach employs KMeans clustering applied directly to the semantic embeddings generated by the transformer model. Unlike LDA, which depends on word frequency, KMeans operates in a continuous embedding space, grouping job descriptions based on their semantic similarity. The algorithm partitioned the embedding space into six clusters, minimizing intra-cluster variance and maximizing inter-cluster separation.

Each cluster is subsequently analyzed using vectorization to extract the most representative terms, providing a bridge between geometric (embedding-based) and lexical (word-based) perspectives. This procedure enables the identification of conceptually cohesive groups such as marketing and personalization, data analytics, management and innovation. Furthermore, word clouds are generated for each cluster to visualize their linguistic markers, paralleling the visualization approach used in LDA.

KMeans proves effective for detecting broad thematic structures and conceptual proximities across job roles, offering an interpretable geometric model that complements the probabilistic and transformer-based methods.

Formally, given a set of $n$ document embeddings $\{e_1, e_2, \dots, e_n\} \in \mathbb{R}^d$, K-Means aims to find $K$centroids $\{\mu_1, \dots, \mu_K\}$, that minimize the within-cluster sum of squared distances (WCSS):

$$J = \sum_{i=1}^{K} \sum_{e_j \in C_i} \parallel e_j - \mu_i \parallel^2 \tag{12}$$

The centroid of each cluster $C_i$ is iteratively updated as the mean of all points assigned to it:

$$\mu_i = \frac{1}{|C_i|} \sum_{e_j \in C_i} e_j \tag{13}$$

Each embedding vector is reassigned to the nearest cluster according to the Euclidean distance:

$$\text{assign}(e_j) = \arg\min_i \parallel e_j - \mu_i \parallel^2 \tag{14}$$

The algorithm iterates between eq. (13) and eq. (14) until convergence, defined as no further change in cluster assignments or a minimal improvement in the objective function $J$.

After clustering, the content of each group is vectorized using TF-IDF weighting to extract the most representative lexical units. For each cluster $C_i$, the average TF-IDF weight of term $t$ is given by:

$$w_{i,t} = \frac{1}{|C_i|} \sum_{d \in C_i} \text{tfidf}_{d,t} \tag{15}$$

**3.2.5. Temporal forecasting, correlation and sectoral analysis**

Following the extraction and normalization of skill mentions from the cleaned corpus, the next analytical stage focuses on modeling temporal evolution, inter-category relationships and sectoral diffusion patterns between 2018 and 2025. Three complementary modules are implemented to address these dimensions: temporal forecasting, correlation analysis and sectoral decomposition.

**3.2.5.1. Temporal forecasting analysis**

The forecasting module aims to project the evolution of demand for distinct skill categories over time, thereby identifying medium-term trends in labor market adaptation to AI technologies. To this end, annualized frequencies of detected skill mentions are computed for five standardized categories: AI_Data, Routine, Soft_Meta, Domain_Specific, and Leadership. Each time series is normalized per 1,000 job postings to ensure comparability across years and data sources. Forecasting is implemented using autoregressive integrated moving average (ARIMA) models combined with exponential smoothing, reflecting both long-term trends and short-term oscillations. Two configurations are applied for robustness: a smoothed ARIMA(1,1,1) model for general trend estimation, and an ARIMA(2,0,2) specification to capture cyclical or volatile dynamics. The models are trained and validated for each skill category independently, generating forecasted values for subsequent years.

**3.2.5.2. Correlation analysis**

The correlation module extended the temporal modeling by investigating interdependencies among skill categories. Using the normalized annual frequencies generated in the previous step, Pearson correlation coefficients are computed pairwise across all skill families. It produces a symmetric correlation matrix, which is visualized through a heatmap, facilitating the identification of co-evolution patterns between categories such as technical, routine and managerial skills. The analysis enables the detection of both positive and inverse relationships, offering a statistical perspective on complementarity and substitution effects among skills. By linking skill trajectories through correlation coefficients, this module provides an empirical measure of convergence and divergence between human-centric and AI-centric competencies.

### 3.2.5.3. Sectoral analysis

This Python module introduces an additional dimension of granularity through sectoral decomposition. Job postings are classified into nine macro-sectors: IT, Healthcare, Legal, Education, Design, Finance, Logistics, Sales and Management, based on regular expression keyword matching within job descriptions. This rule-based classification system is optimized for recall and validated on sampled subsets to ensure accurate sector identification. Following classification, the skill detection pipeline is re-applied within each sector to compute normalized mention rates, both cumulatively and over time.
It illustrates the diffusion of AI-related terminology across industries. Methodologically, the sectoral analysis integrated the temporal and correlation outputs, revealing how skill categories evolve heterogeneously across domains. By comparing normalized trajectories per sector, the module enabled the quantification of domain-specific adoption patterns and the identification of temporal asymmetries in AI diffusion.

## 3.3. Temporal alignment and normalization

To analyze longitudinal dynamics, all postings are aligned into yearly intervals covering 2018-2025. For each category and sector, skill mentions are normalized by the total number of postings in that year, producing standardized metrics expressed as mentions per 1,000 job postings. The temporal modeling and forecasting stages are implemented using several Python modules which applied time-series models such as exponential smoothing and ARIMA to capture both smooth and oscillatory skill trajectories.

# 4. Results

The results of the analytical pipeline highlight several trends. Since 2021, there has been a sharp increase in AI-related skill mentions, coinciding with the diffusion of LLMs and generative tools. This growth is strongest in technology-intensive fields, particularly IT, Finance and e-commerce, where automation and personalization drive performance. At the same time, the decline in Routine skills reflects a transition toward automation, while framing patterns indicate a hybrid dynamic in which AI serves as an augmentative partner rather than a substitute. Positive correlations between AI_Data and Leadership suggest that strategic roles are becoming more data-driven, whereas Routine and AI_Data are inversely related, confirming the reduction of repetitive tasks. In e-commerce, this transformation is most evident: new hybrid roles such as *AI Marketing Analyst* or *Product Data Strategist* illustrate how generative-AI enhances analytical and creative processes. Employers frame AI as a productivity amplifier, not a replacement for human expertise. Forecasting models further indicate that AI_Data and Soft_Meta competencies will continue to expand through 2025, while Routine skills will keep declining, signaling a lasting shift toward data-centric, ethically guided and strategically integrated forms of digital work.

## 4.1. Skill extraction and categorization

Figure 2 illustrates the longitudinal evolution of skill mentions per 1,000 job postings, based on the standardized lexical categories. A visible drop occurs in 2021 across all categories due to the reduced data availability during the pandemic, which disrupted recruitment cycles. After this temporary dip, AI_Data, Soft_Meta and Domain_Specific skills show a sharp recovery in 2022, indicating renewed emphasis on digital and analytical competencies following the global acceleration of AI adoption.

Domain_Specific and Soft_Meta skills consistently maintain the highest frequencies, reflecting the sustained importance of specialized and transversal abilities. Leadership follows a similar but more moderate trend, while Routine skills remain relatively stable at lower levels. Figure 2 captures a structural shift in labor demand: post-pandemic recovery coincides with the rise of AI-driven, cognitive and domain-specific competencies, signaling a transition toward more analytical and technology-integrated job profiles.

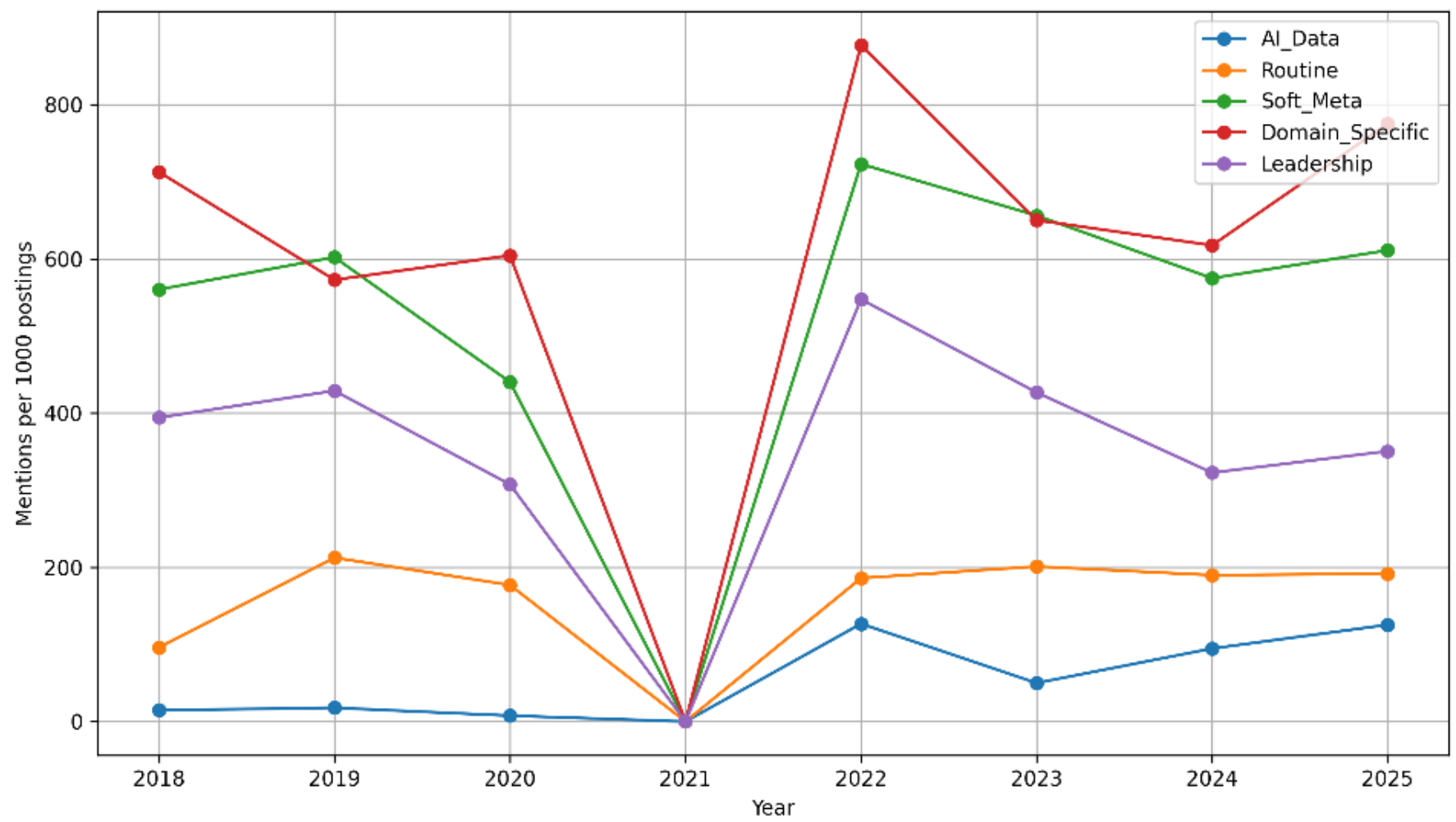

Figure 2. Skill mentions per 1000 job postings

### 4.2. Semantic-shift and framing analysis

As shown in the left plot in Figure 3, the FI displays a sharp peak in 2021, reflecting an intensified perception of AI as an augmentative rather than substitutive force. This peak coincides with the post-pandemic recovery period, when organizations accelerated digital transformation and emphasized human-AI collaboration to sustain hybrid and remote work structures. The subsequent decline between 2022 and 2023 suggests a temporary shift toward automation-focused narratives, likely driven by efficiency pressures and the rapid diffusion of generative models such as GPT-3 and early ChatGPT prototypes. However, the second right plot in Figure 3 indicates that, despite short-term oscillations, the semantic proximity to *AI* and *Augment* anchors remains consistently higher than to *Automate* terms. This confirms a stable discursive orientation toward augmentation, AI as a supportive, knowledge-enhancing instrument rather than a labor-replacing mechanism. The mild decline in the augmentative framing after 2023 may be interpreted as a phase of normalization and critical reassessment of generative-AI systems, reflecting both ethical concerns and uncertainty about long-term integration. By 2024-2025, the modest recovery in the FI corresponds to the institutional stabilization of AI adoption across industries. Organizations increasingly frame AI as a co-creative and decision-support partner, signaling a maturing perception of hybrid intelligence rather than technological substitution. This transition suggests that the discourse surrounding AI in job postings has evolved from experimental enthusiasm toward pragmatic, governance-oriented integration.

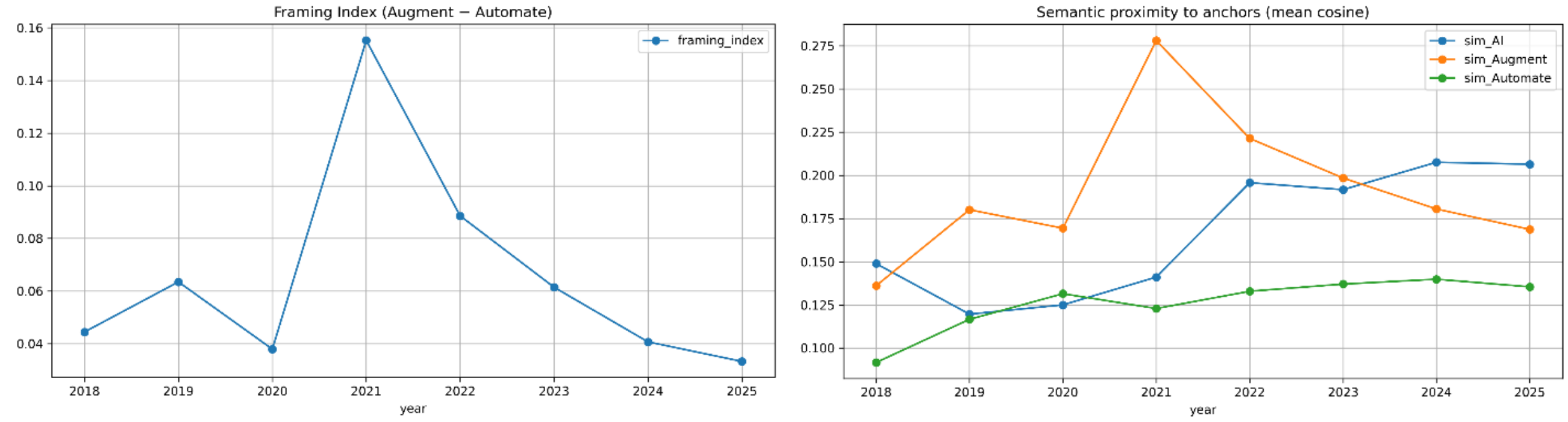

Figure 3. Temporal dynamics of AI-related semantic proximity and framing index (2018-2025)

### 4.3. Sectoral analysis

The exceptionally high concentration of domain-specific terminology in *Healthcare* indicates that automation and AI systems have not yet penetrated deeply into clinical or diagnostic roles (as in Figure 4), where regulatory, ethical and safety constraints continue to favor human oversight. In contrast, the relatively lower weight of AI_Data and Leadership skills in Healthcare points to a slow digital transition, limited primarily to data management and administrative optimization rather than core medical decision-making. In *IT*, the pattern is almost the opposite. Notably, AI_Data and Soft_Meta skills dominate, marking this sector as the epicenter of generative-AI diffusion. The strong presence of transversal competencies such as communication, problem-solving and adaptability reflects the hybridization of technical and managerial profiles, professionals are expected not only to build AI systems but also to integrate them ethically and strategically into production environments. This combination signals a maturing AI ecosystem, where technical innovation is increasingly paired with governance and leadership functions. The *Legal* sector exhibits a similar dual structure, with high Domain_Specific and Soft_Meta scores, suggesting early-stage digital transformation mediated by language-based automation tools (like document summarization, compliance screening). However, the relatively modest presence of AI_Data skills implies that legal professionals remain consumers rather than developers of AI solutions, relying on external technologies rather than in-house model expertise. In *Finance*, the balance between Soft_Meta, Domain_Specific and Leadership categories suggest an environment of controlled experimentation. Financial institutions have embraced AI for analytics and risk modeling but continue to emphasize governance and interpretability, resulting in a skill mix that values analytical literacy and ethical oversight rather than deep model engineering. The *Education* sector shows an atypical trajectory. While AI_Data mentions are generally modest compared to IT or Finance, there is a clear spike around 2022, indicating a short-lived but intense phase of experimentation with AI-related competencies. This surge likely corresponds to the post-pandemic acceleration of digital transformation in learning environments, when educational institutions rapidly adopted AI-powered tools for online assessment, adaptive learning and content generation. *Design and Sales* occupy an intermediate zone where AI serves primarily as an augmentation tool. In these creative and customer-oriented domains, the relatively balanced mix of Soft_Meta, Routine and Leadership competencies indicates the integration of AI-assisted design, marketing automation and personalization technologies. Notably, AI acts as a force multiplier for human creativity rather than a substitute, amplifying productivity while preserving the centrality of human judgment.

These models reveal a pronounced upward trend in AI_Data skills and a corresponding decline in Routine tasks, reflecting the progressive automation of repetitive roles through cognitive systems and generative-AI technologies. Particularly in e-commerce, the frequency of AI-related skill mentions rose sharply after 2022, aligning with the widespread commercial adoption of recommendation engines, generative marketing tools and automated analytics platforms.

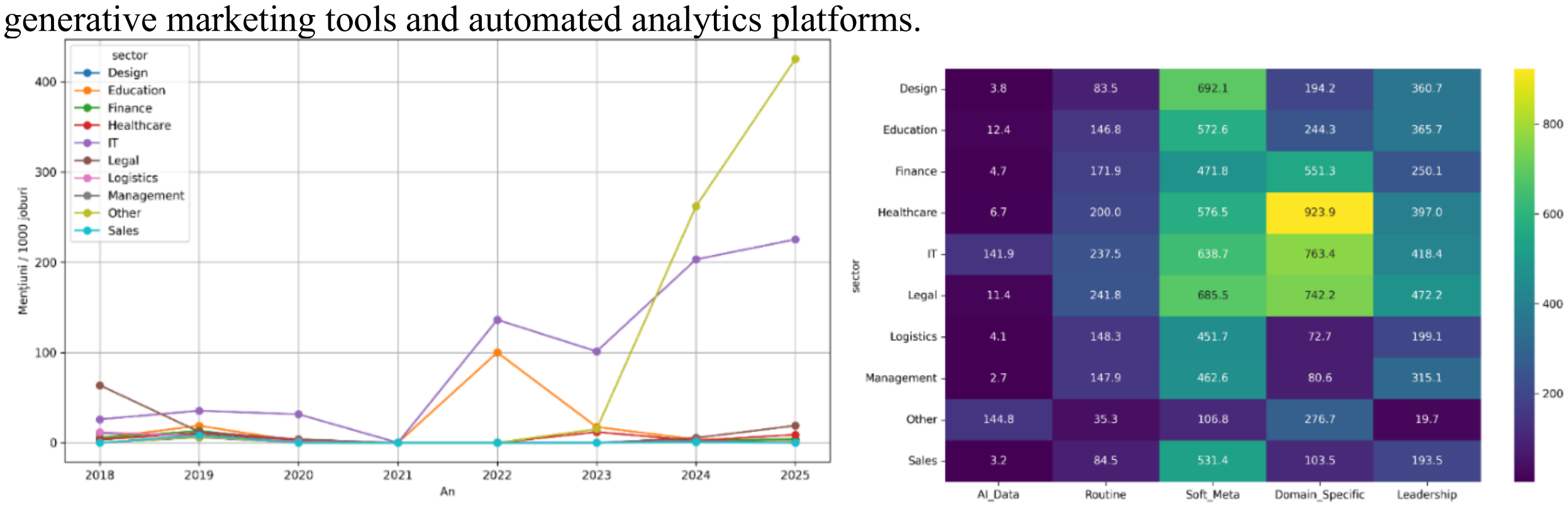

Figure 4. Evolution of AI mentions across sectors and corresponding skill profiles

### 4.4. Correlation analysis

The correlation matrix in Figure 5 illustrates the interdependencies between the five skill dimensions: *AI_Data*, *Routine*, *Soft_Meta*, *Domain_Specific* and *Leadership*. Pearson coefficients range between 0.49 and 1.00, indicating varying degrees of association among categories. The highest correlations appear between *Soft_Meta*, *Domain_Specific* and *Leadership* (above 0.9), forming a cohesive cluster of cognitive, professional and managerial competencies. These categories frequently co-occur within job postings, suggesting that analytical reasoning, domain expertise and strategic oversight are increasingly integrated within the same roles. In contrast, *AI_Data* exhibits the weakest correlations with other skill groups, ranging from 0.49 to 0.63. This relative independence indicates that AI- and data-related skills often appear in more specialized or technically focused postings, distinct from those emphasizing human-centric or managerial capacities. *Routine* skills occupy an intermediate position, maintaining moderate correlations with both *AI_Data* and the human-managerial cluster, reflecting a transitional space where automation and oversight functions coexist.

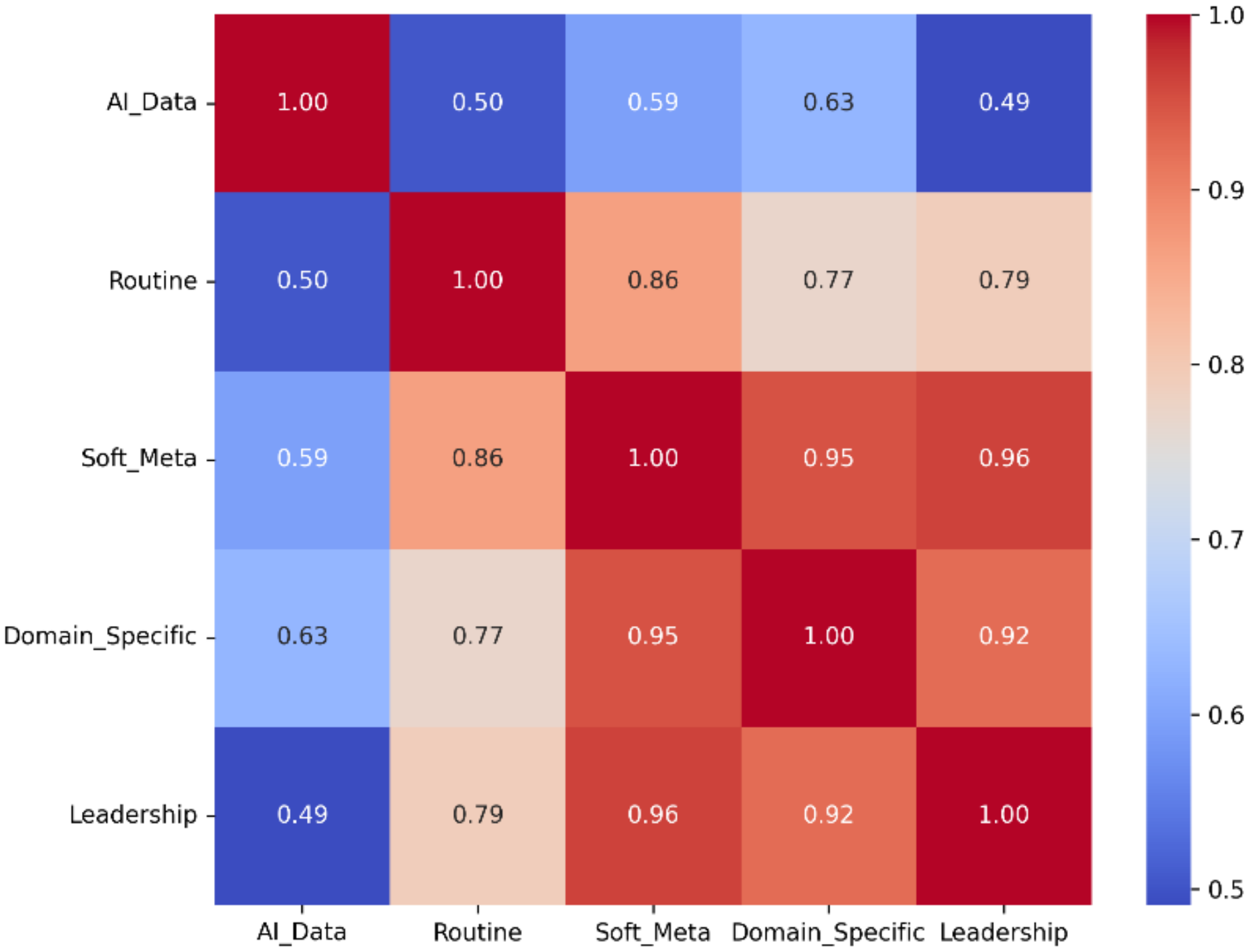

Figure 5. Correlation matrix

## 4.5. Semantic topic modeling with BERTopic, LDA and KMeans

### 4.5.1. BERTopic analysis

Figure 6 depicts the evolution of significant themes over time extracted from the BERTopic output. It captures the gradual transformation of the digital landscape and the shifting demand for competencies across sectors associated with e-commerce, finance and technology-driven business models. Between 2018-2020, the trend shows an early phase dominated by themes such as Corporate/Manufacturing and Accounting/Receivables, reflecting the first wave of process digitalization and the integration of data-based management systems. The COVID-19 pandemic amplified these tendencies, leading to a sharp rise in Finance and Banking topics, which signals the migration of financial operations toward fully digital environments and the growing reliance on automated infrastructures to sustain the surge in online transactions.

After 2020, dynamics reveal a deeper structural shift, from basic digitalization to human-machine augmentation through predictive and analytical AI. The Finance topic exhibits the steepest increase between 2021-2024, suggesting that the financial sector, particularly within e-commerce ecosystems, became a pioneer in adopting AI-powered tools for data analytics, decision automation and commercial optimization. In parallel, the rise of Tech Business points to the globalization of this transformation: AI is no longer a background technology but a strategic driver reshaping business models, logistics and market personalization at scale. Meanwhile, Corporate/Manufacturing follows a more oscillating path, partly due to the pandemic's impact on supply chains and production but also because of a gradual transition toward hybrid forms of automation and augmentation. This trend is consistent with the FI analysis, which documented a semantic shift from viewing AI as a substitute for human labor ("automate") to positioning it as a collaborative and supportive partner ("augment"). Similarly, Accounting/Receivables and Banking reflect the consolidation of digital maturity within financial operations, where human roles increasingly emphasize analytical interpretation, ethical oversight and strategic decision-making, while repetitive tasks are progressively automated.

A particularly notable moment occurs between 2023-2024, coinciding with the mainstream diffusion of generative-AI models, such as ChatGPT. This period marks a convergence between technological and commercial domains, with a surge in AI adoption accompanied by growing awareness of its strategic, ethical and interpretative implications. Employers begin to emphasize AI literacy, the ability to understand, interpret and responsibly apply model outputs, as a defining competency in digital commerce. By 2025, the relative decline in topic frequencies indicates a phase of market consolidation rather than regression. Organizations appear to be moving from the experimental enthusiasm that characterized the initial adoption of AI tools to a more deliberate and mature integration, where AI becomes a stable component of analytical and decision-support systems.

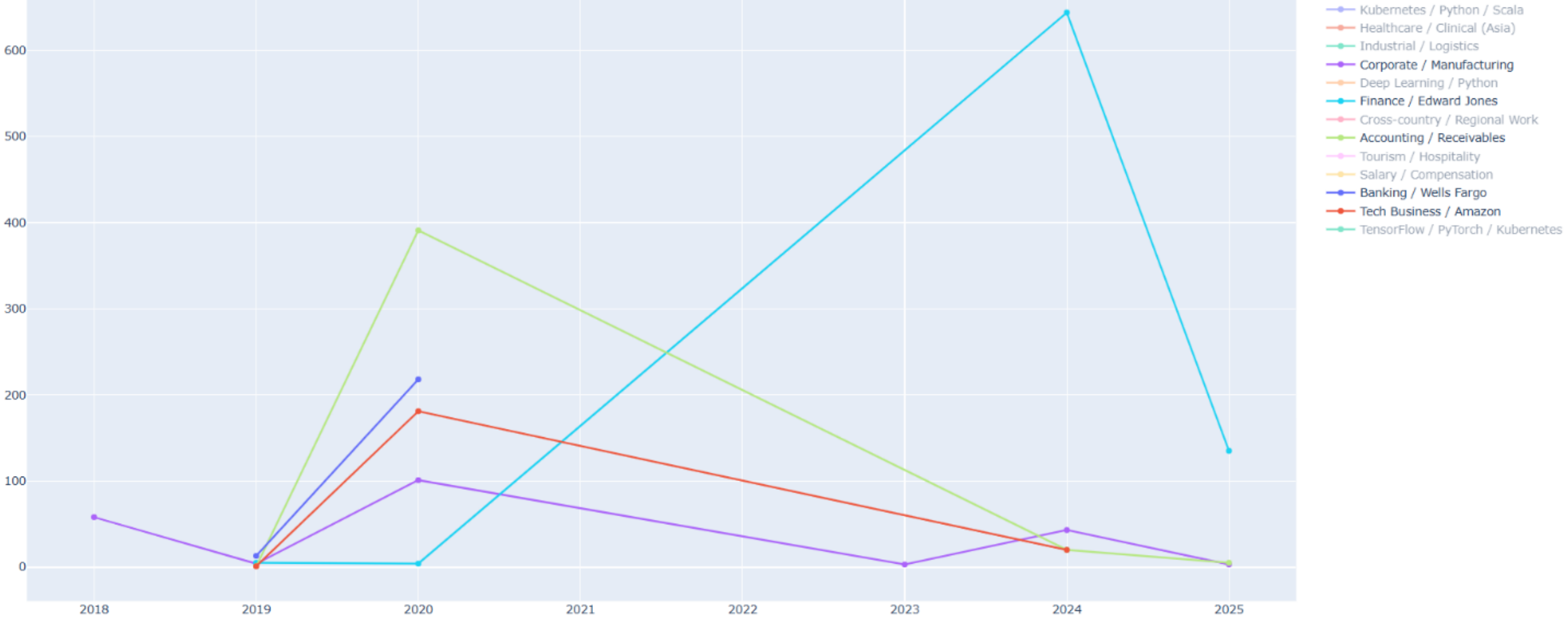

Figure 6. Evolution of themes over time with BERTopic

Each topic in Figure 7 encapsulates a distinct cluster of labor-market discourse, organized around core skill and activity lexicons, thereby illustrating the latent semantic segmentation of the job landscape. Topic 0 represents the marketing and commercial communication domain, centered on terms such as *marketing*, *sales* and *customer*. This lexical field captures the language of e-commerce, CRM and digital promotion, where AI is increasingly integrated into automation tools for campaigns, behavioral analytics and customer personalization. It illustrates an augmentative pattern, in which AI amplifies creative and operational efficiency rather than replacing human agency in market interaction. Topic 1 revolves around *care* and *patient*, delineating the healthcare and caregiving sector. The vocabulary emphasizes interpersonal and ethical dimensions, reflecting the human-centered nature of medical practice. In this area, AI is framed as a supportive technology for diagnostics, monitoring and patient management, functioning as an analytical assistant while preserving professional judgment. Automation remains constrained by regulatory and moral considerations. Topic 2 corresponds to the AI and data science domain, structured around highly technical terms such as *python*, *tensorflow*, *pytorch*, *mathematics*, *kubernetes*, *hadoop* and *statistics*. This topic encapsulates the linguistic core of computational infrastructure, data engineering and machine learning specialization. The strong co-occurrence of programming frameworks and orchestration tools suggests that model training pipelines and cloud-based automation form the backbone of AI-related professional discourse, underscoring persistent technical specialization.

Topic 3 focuses on *accounting*, *financial*, *accounts*, *finance* and *tax*, representing the financial and administrative sector. Its formal and standardized vocabulary signals a mature discourse shaped by precision, compliance and auditability. Within this field, AI adoption is procedural and explainable, used primarily for risk modeling, anomaly detection and reporting automation, emphasizing reliability over disruption. Topic 4 brings together *clinical*, *medical*, *pharmaceutical* and *regulatory*, corresponding to the medical-pharmaceutical research ecosystem. Compared to Topic 1, the language is more institutional and compliance-oriented, reflecting the role of AI in regulated experimentation and data-driven medicine. AI is conceptualized as an augmentative analytical instrument, enhancing precision, discovery and efficiency in data interpretation rather than displacing human expertise. Topic 5 features terms such as *hr*, *payroll*, *employee*, *recruitment* and *resources*, mapping to the human resources and organizational management domain. The lexical focus on personnel management and evaluation points to the diffusion of AI into workplace analytics. Automation streamlines repetitive and administrative tasks, while interpretive and social dimensions remain human-led, maintaining a supportive, augmentative role for AI in people management.

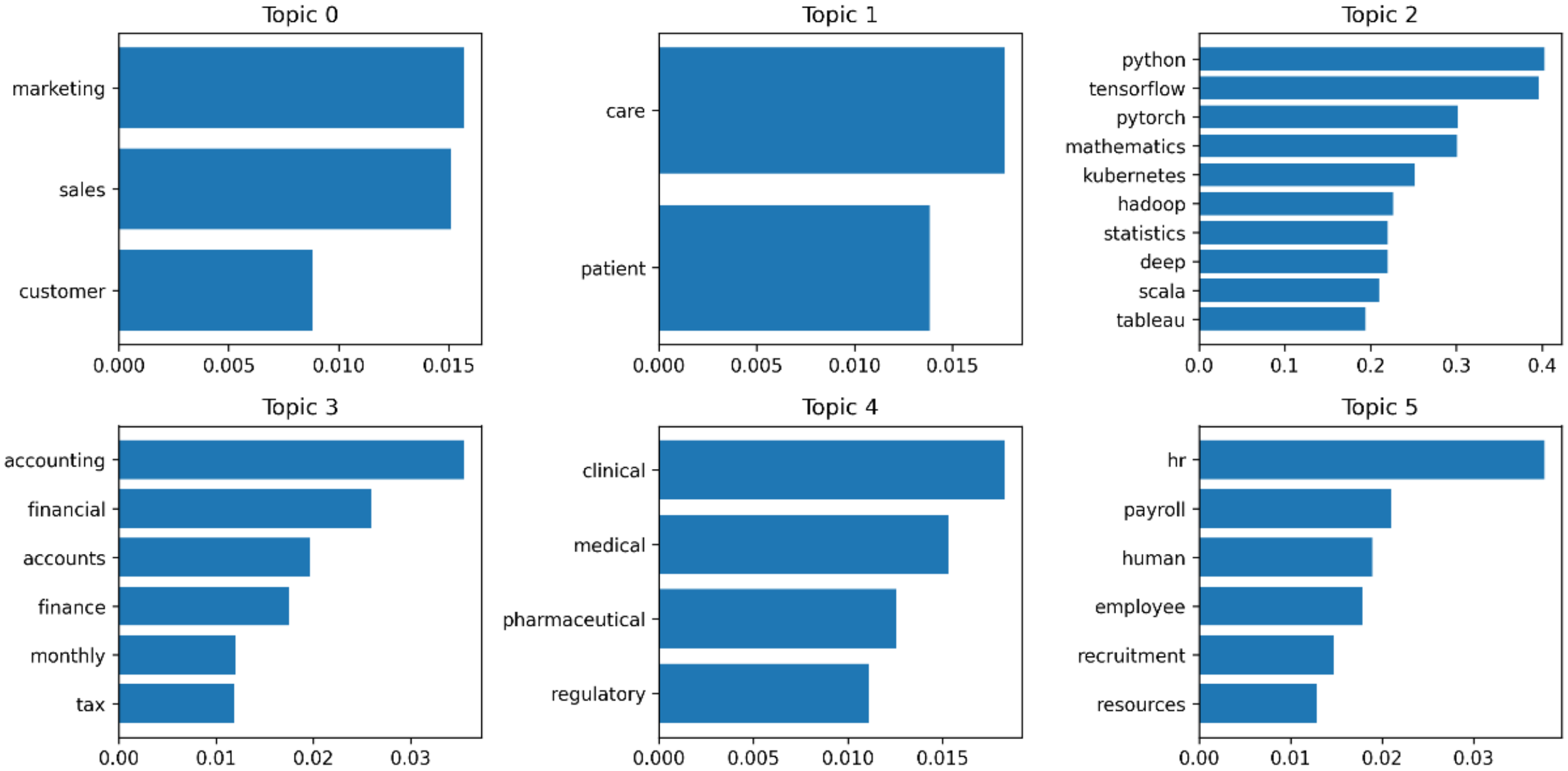

Figure 7. BERTopic generated topics

**4.5.2. LDA analysis**

The first topic obtained with LDA (Figure 8) is characterized by words such as *customer*, *apply*, *role*, *skills*, *work* and *company*, which indicate a focus on customer-facing and service-oriented occupations. This topic reflects the operational and interpersonal dimension of AI adoption, where roles emphasize adaptability, collaboration and skill diversification. The second topic, dominated by terms like *technical*, *support*, *development*, *data* and *project*, reflects the technical infrastructure underlying AI deployment. It emphasizes the demand for professionals capable of integrating data-driven systems, maintaining automated workflows and providing technical support.

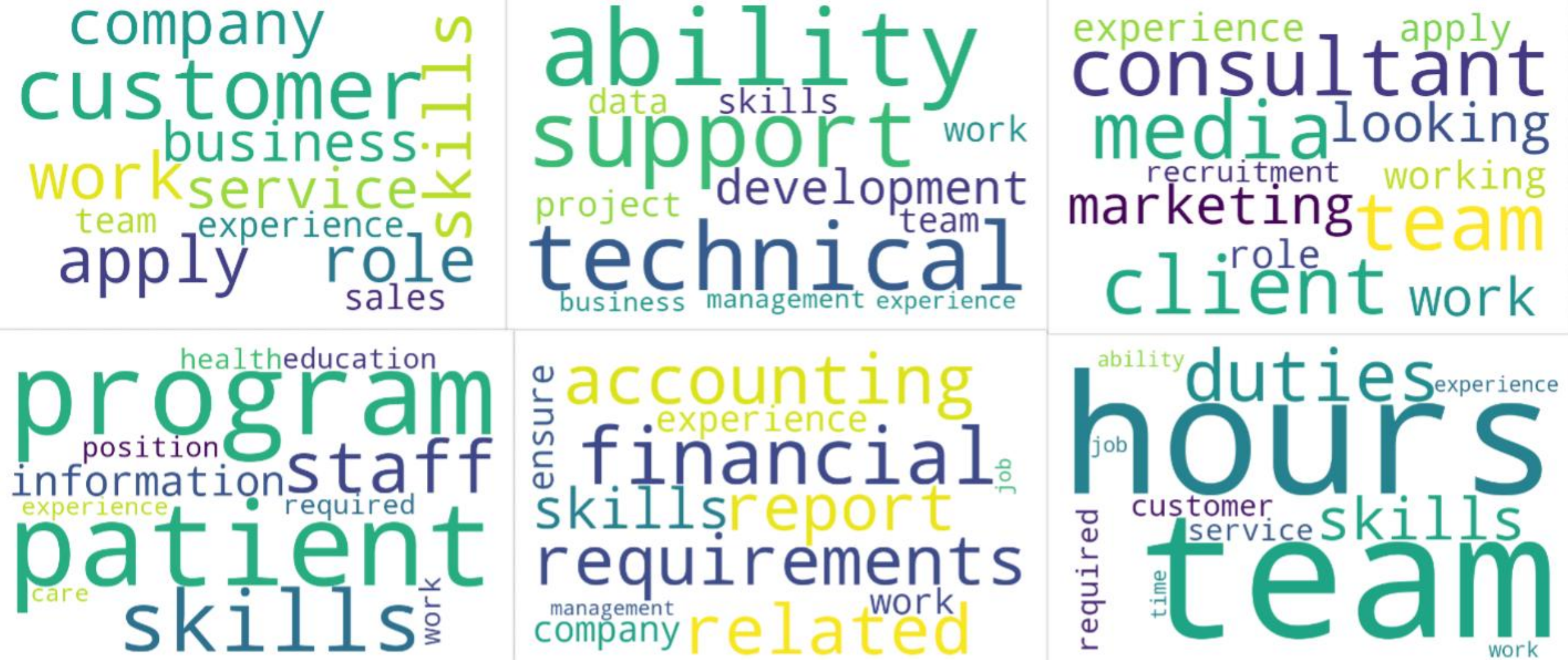

Figure 8. Wordclouds-LDA

In the third topic, words such as *consultant*, *client*, *marketing*, *media* and *team* reveal an orientation toward digital marketing, creative consultancy and AI-assisted communication strategies. This cluster highlights how AI tools are embedded in marketing ecosystems, reshaping recruitment, branding and customer engagement through personalized and data-driven media practices. The fourth topic, represented by *patient*, *program*, *staff*, *education* and *health*, indicates an association with healthcare and educational transformation. The fifth topic revolves around *financial*, *accounting*, *report*, *requirements* and *management*, pointing to the financial and analytical dimension of AI applications. Finally, the sixth topic, featuring words such as *team*, *hours*, *duties*, *service* and *skills*, captures the operational and organizational routines associated with modern work structures.

**4.5.3 KMeans analysis**

The first cluster obtained with KMeans (as in Figure 9) corresponds to the financial and accounting sector, where the dominant words, *accounting*, *financial*, *experience* and *sales*, indicate roles focused on business operations, reporting and corporate management. The association between *sales* and *financial* terms shows the growing overlap between financial analysis and commercial strategy, suggesting that economic roles are increasingly data-driven and connected to e-commerce performance indicators. The second cluster reflects the healthcare and clinical field, characterized by the frequent appearance of *patient*, *medical*, *clinical* and *care*. It represents the ongoing digitalization of healthcare, where data processing, monitoring systems and AI-based diagnostics support medical decision-making. The third cluster captures the technical and programming area, defined by *python*, *tensorflow*, *pytorch*, *kubernetes* and *sql*. These terms point to positions related to software development, data infrastructure and machine learning model deployment. The fourth cluster includes terms like *management*, *project*, *team*, *data* and *experience*, describing project coordination and organizational management. This cluster emphasizes the growing need for professionals who can translate technical innovation into operational efficiency. The fifth cluster is centered on recruitment, communication and client relations. Words such as *recruitment*, *client*, *apply* and *role* reveal a focus on human resource management and personalized engagement. The sixth cluster groups together the most service-oriented terms, *customer*, *service*, *skills*, *job* and *work*, illustrating the direct interface between businesses and consumers. This vocabulary is closely associated with e-commerce, where companies rely on AI systems to manage customer interactions, predict behavior and tailor user experiences. The repetition of *customer* and *service* emphasizes the centrality of personalization and responsiveness in the digital economy, as companies strive to balance automation with human-centered design.

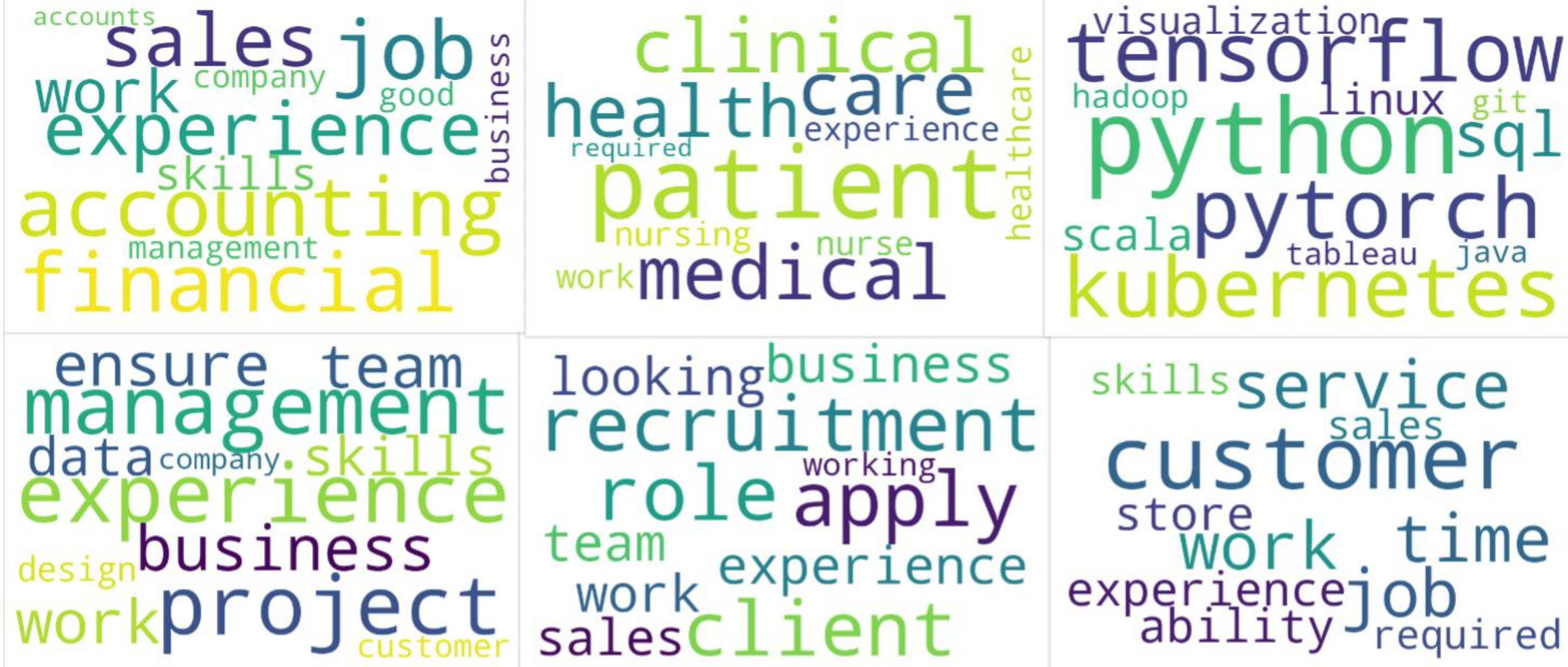

Figure 9. Wordclouds LDA

**4.5.4. Comparison of the results**

A comparison of the results obtained by the three techniques is offered in Table 5.

Table 5. Thematic correspondence across techniques

| Core theme | BERTopic (Transformer-based) | LDA (Probabilistic) | KMeans (Embedding-based) |
|---|---|---|---|
| Organizational and workforce transformation | AI Governance, Workforce AI Adoption | customer, company, service, role, skills | operations, management, leadership, client |
| Education and workforce preparation | Learning Adaptation, AI Training | patient, staff, program, education | training, support, communication, learning |
| Technical and data foundations | Model Development, LLM Engineering, MLOps | technical, data, engineer, project, development | analytics, python, automation, systems |
| E-commerce and sales innovation | E-commerce Automation, Generative Content Marketing | marketing, content, media, sales | campaign, customer, digital, performance, recommendation |
| Societal and creative implications | Creative Design Augmentation, Ethical AI | consultant, creativity, ethics, innovation | communication, design, ethics, creativity |

**4.6. Cross-sectoral framing trends**

The sectoral framing analysis in Figure 10 illustrates the heterogeneous ways in which AI and automation are being linguistically positioned across professional domains. By comparing the trajectories of AI, Augment and Automate framings between 2018-2025, a divergence emerges in how sectors conceptualize the integration of intelligent technologies, ranging from cautious experimentation to structural adoption and normalization. In the *IT* sector, the framing of AI shows a consistent upward trend, consolidating after 2021. The discourse is characterized by a pragmatic balance between innovation and control. Mentions of augmentation rise sharply in 2022, corresponding to the diffusion of generative-AI and automation frameworks into software development, data engineering and DevOps environments. The language of augmentation dominates over that of automation, reflecting a shift from viewing AI as a threat to technical roles toward understanding it as an efficiency multiplier that enhances coding, testing and deployment workflows.

The *management* sector presents a similar evolution but with a stronger and earlier adoption of augmentation-oriented framing. This indicates that organizational discourse increasingly centers on AI-assisted decision-making, strategic foresight and leadership in digital transformation. In *finance*, the narrative is shaped by a dual emphasis on automation and augmentation. Early fluctuations give way to a steady increase in augmentation framing after 2022, coinciding with the expansion of AI in risk modeling, fraud detection and predictive analytics. However, automation retains a secondary presence, revealing the sector's reliance on rule-based systems and its regulatory caution. The *education* sector displays a distinctive pattern, with a pronounced spike in AI-related framing in 2022. This anomaly corresponds with the mass adoption of EdTech and learning analytics tools during the post-pandemic digital shift. However, the trend stabilizes afterward, suggesting that AI integration in education remains situational rather than systemic. The dominance of augmentation language underscores the framing of AI as a pedagogical assistant rather than a substitute for educators, enhancing assessment, adaptive learning and content personalization while preserving human oversight and empathy as core values of the teaching profession.

In *healthcare*, the discourse remains grounded in caution. Domain-specific and ethical constraints limit the prevalence of automation framing, while augmentation dominates. This reflects the sector's emphasis on AI-assisted diagnostics, patient monitoring and administrative optimization, rather than fully autonomous systems. The pattern points to a narrative of collaborative intelligence, where AI augments clinical judgment but does not replace it. The temporary rise in AI framing in 2022-2023 may correspond to the pandemic's acceleration of telemedicine and medical imaging analytics. The *legal* sector mirrors healthcare in its conservatism, showing modest growth in AI and augmentation framings but minimal automation language. The *logistics* sector exhibits volatility, with short-lived peaks in AI and augmentation framings around 2022. This likely corresponds to post-pandemic restructuring of global supply chains and the integration of predictive analytics in routing and inventory systems.

In *sales* and particularly in *e-commerce*, the discourse undergoes a striking transformation. After 2021, augmentation framing rises dramatically, accompanied by sustained growth in AI mentions. This linguistic pattern reflects the mainstream adoption of AI-driven recommendation engines, marketing automation and content generation. E-commerce stands out as the sector where AI is most tightly coupled with creativity and personalization. The coexistence of high augmentation and moderate automation framing signals a redefinition of human-machine interaction: automation handles transactional efficiency, while humans leverage AI to craft narratives, optimize engagement and innovate in customer experience. The language reflects co-creativity rather than replacement, making e-commerce a prototype for future hybrid digital labor models.

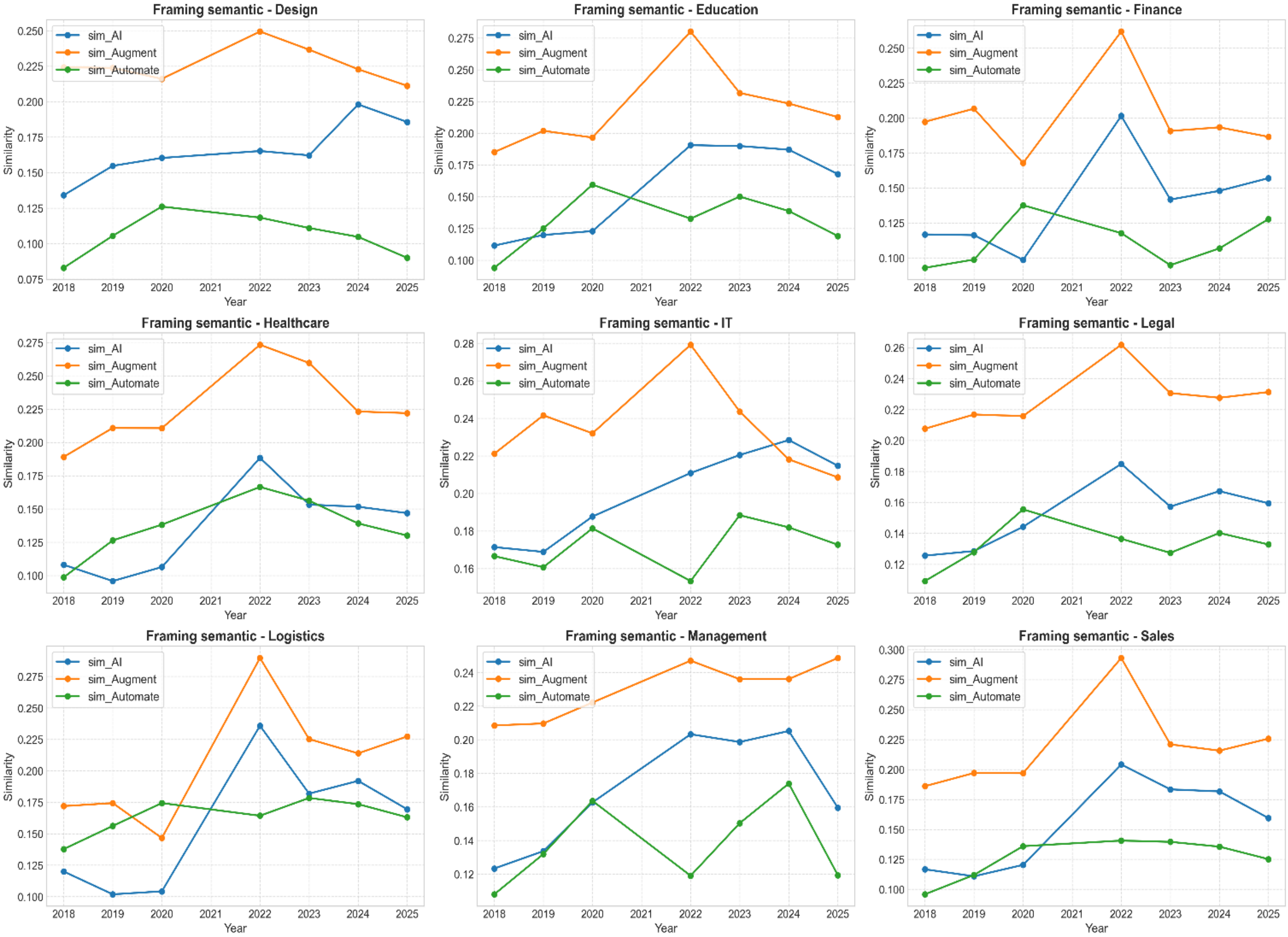

Figure 10. Semantic framing across sectors

Findings showed a consistent trend toward augmentation-oriented framing, especially after 2021, when the adoption of generative-AI technologies accelerated. In e-commerce, this tendency was particularly pronounced, with frequent references such as AI-assisted marketing, decision-support analytics, and human-in-the-loop optimization. These expressions framed AI as a co-creative and productivity-enhancing force, reinforcing human roles in strategic and creative functions. Conversely, the IT and Finance sectors leaned more toward automation-oriented language, reflecting widespread process optimization. In contrast, Healthcare, Design, and Education maintained predominantly collaborative framings.

**4.7. Forecast**

The comparative ARIMA forecasts in Figure 11 provide complementary perspectives on the temporal dynamics of skill demand across the five macro-categories. The smoothed ARIMA(1,1,1) projection models a stable, gradually converging trend after 2025, indicating that the market for digital and cognitive competencies is approaching a structural equilibrium. The trajectories of Domain_Specific and Soft_Meta skills remain dominant, showing steady growth that plateaus at high frequency levels, reflecting their consolidation as fundamental components of employability in AI-mediated labor markets. Leadership follows a parallel but lower trajectory, suggesting a sustained managerial demand oriented toward coordination and oversight rather than expansion. Conversely, Routine and AI_Data exhibit mild declines in the smoothed scenario, consistent with a gradual automation of repetitive tasks and the standardization of technical AI competencies. The smoothing effect implies a stabilization of demand rather than volatility, suggesting that the diffusion of AI tools has reached maturity and is now embedded in broader professional frameworks.

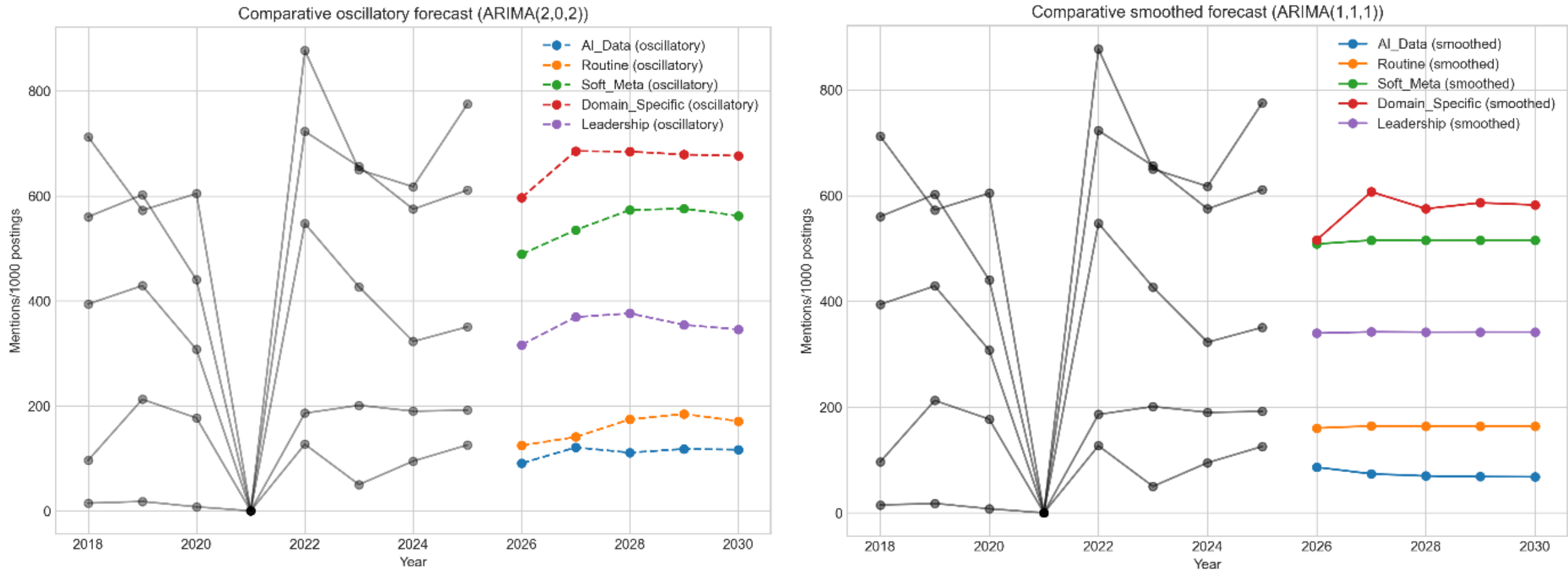

Figure 11. Forecast

The oscillatory ARIMA(2,0,2) model captures cyclical fluctuations that the smoothed specification attenuates. Notably, Soft_Meta and Leadership continue to expand moderately, but with visible short-term oscillations, indicating alternating phases of emphasis on adaptive and managerial capabilities. Domain_Specific remains robust yet shows slight periodic corrections, which may reflect the saturation of specialized roles.

## 5. Conclusions

The analysis reveals an accelerating integration of AI-related competencies across professional domains, particularly after 2021, when generative-AI entered mainstream application through LLMs and automated content systems. Mentions of skills such as prompt engineering, data validation and model monitoring have expanded rapidly, especially within IT, Finance and e-commerce, signaling a structural redefinition of what constitutes digital expertise.

At the same time, the steady decline of Routine skill mentions points to an ongoing process of task automation, yet the data suggests that this is not a simple substitution dynamic. Rather than replacing human labor, AI is increasingly portrayed as an augmentative partner, a tool that enhances analytical reasoning, creativity and strategic decision-making. Employers appear to value professionals who can integrate AI tools into existing workflows, interpret their outputs and ensure ethical oversight.

Sectoral variation reinforces this interpretation. Technology and financial fields lead in AI adoption, while e-commerce demonstrates a hybrid pattern that bridges automation and human creativity. Job descriptions in this sector increasingly reference AI-assisted marketing, personalized recommendation systems and data-driven content optimization, reflecting a form of applied human-AI collaboration where automation enhances productivity without displacing creative roles. Education shows a temporary spike in AI-related terminology in 2022. By contrast, healthcare and legal sectors remain grounded in domain-specific terminology, underscoring the cautious and regulated nature of these environments. The next phase of workforce development will thus hinge on cultivating AI fluency, the capacity to guide, question and strategically employ AI in support of human judgment and creativity.

The framing perspective provides insight into how organizations conceptualize the role of AI in their workflows: as a collaborative tool that amplifies human capability (augmentative framing) or as an autonomous system that substitutes human labor (automative framing). Two hypotheses guide this analysis. The first one posits that firms in high-adoption sectors such as IT and e-commerce will shift earlier toward augmentative framing, emphasizing human-AI collaboration and oversight. The second one suggests that, across industries, increases in generative-AI skill mentions will coincide with a decline in automation-oriented framing, reflecting a broader cultural and organizational transition toward viewing AI as an assistive rather than a replacement technology.

While the study adopts a robust and reproducible computational framework, several methodological limitations must be acknowledged. First, job postings primarily reflect stated requirements, which may not accurately correspond to the actual skills used in daily professional activities or the final hiring outcomes. Despite de-duplication procedures, some postings may be reposted or cross-listed across multiple platforms, potentially inflating certain categories. Furthermore, the presence of AI-related terminology can occasionally serve marketing or branding purposes rather than denote genuine technical needs. Second, sectoral imbalances exist due to uneven data availability: digital industries such as IT and e-commerce are heavily represented in online job boards, whereas traditional sectors like education and healthcare appear less frequently.

**Declarations**

**Funding**-This work was supported by a grant of the Ministry of Research, Innovation and Digitization, CNCS/CCCDI-UEFISCDI, project number COFUND-CETP-SMART-LEM-1, within PNCDI IV. This research was funded by CETP, the Clean Energy Transition Partnership under the 2022 CETP joint call for research proposals, co funded by the European Commission (GAN° 101069750) and with the funding organizations detailed on https://cetpartnership.eu/funding-agencies-and-call-modules.

**Acknowledgement**-This work was supported by a grant of the Ministry of Research, Innovation and Digitization, CNCS/CCCDI-UEFISCDI, project number COFUND-CETP-SMART-LEM-1, within PNCDI IV. This research was funded by CETP, the Clean Energy Transition Partnership under the 2022 CETP joint call for research proposals, co funded by the European Commission (GAN° 101069750) and with the funding organizations detailed on https://cetpartnership.eu/funding-agencies-and-call-modules.

**Conflicts of interest/Competing interests**-The authors declare that there is no conflict of interest.
**Ethics approval**-Not applicable
**Consent to participate**-Not applicable
**Consent for publication**-Not applicable
**Availability of data and material**-Data will be available on request.
**Authors' contributions**-D.M.P., S.V.O: Conceptualization, Methodology, Formal analysis, Investigation, Writing–Original Draft, Writing–Review and Editing, Visualization, Project administration. A.B, D.M.P: Validation, Formal analysis, Investigation, Resources, Data Curation, Writing–Original Draft, Writing–Review and Editing, Visualization, Supervision.